\documentclass[useAMS]{mn2e}
\usepackage{epsfig,lscape}
\usepackage{graphicx}

\DeclareGraphicsExtensions{.eps,.eps.gz,.epsi}

\newcommand\ao{Appl. Opt.}%
\newcommand\procspie{Proc. SPIE}%
\newcommand\aj{AJ}%
\newcommand\apj{ApJ}%
\newcommand\apjl{ApJL}%
\newcommand\aap{A\&A}%
\newcommand\mnras{MNRAS}%
\newcommand\araa{ARA\&A}%
\newcommand\apjs{ApJS}%
\newcommand\aaps{A\&AS}%

\newcommand{\teff}{\mbox{$T_{\rm eff}$}}
\newcommand{\logg}{\rm{log}~$g$}
\newcommand{\feh}{{\rm [Fe/H]}}
\newcommand{\ebv}{$E_{B-V}$}
\newcommand{\kms}{km s$^{-1}$}

\begin{document}

\title[Detections of DIBs in the SDSS Spectra]{Detections of Diffuse Interstellar Bands in the SDSS Low-resolution Spectra}

\author[Yuan \& Liu]{H.B. Yuan$^{1,2}$ \& X.W. Liu$^{1,3}$\thanks{E-mail: x.liu@pku.edu.cn}\\
\\ 
$1$ Kavli Institute for Astronomy and Astrophysics, Peking University, Beijing 100871, China \\
$2$ LAMOST Fellow \\
$3$ Department of Astronomy, Peking University, Beijing 100871, China}

\date{Received:}

\pagerange{\pageref{firstpage}--\pageref{lastpage}} \pubyear{2012}

\maketitle

\label{firstpage}

\begin{abstract}{
Diffuse interstellar bands (DIBs) have been discovered for almost a century,
but their nature remains one of the most 
challenging problems in astronomical spectroscopy. 
Most recent work to identify and investigate the properties and carriers of
DIBs concentrates on high-resolution spectroscopy of selected sight-lines.
In this paper, we report detections of DIBs in the Sloan Digital Sky Survey
(SDSS) low-resolution spectra of a large sample of Galactic stars.
Using a template subtraction method, we have successfully identified the
DIBs~$\lambda$$\lambda$5780, 6283 in the SDSS spectra of
a sample of about 2,000 stars and measured their strengths and radial
velocities.
The sample is by far the largest ever assembled.
The targets span a large range of reddening, \ebv~$\sim$~0.2 –- 1.0, and are
distributed over a large sky area and involve a wide range of stellar
parameters (
effective temperature, surface gravity and metallicity), confirming that the
carriers of DIBs are ubiquitous in the diffuse interstellar medium (ISM).
The sample is used to investigate relations between strengths of DIBs and
magnitudes of line-of-sight extinction, yielding results (i.e., $EW$(5780)
$= 0.61~\times$~\ebv~
and $EW$(6283) = 1.26~$\times$~\ebv) consistent with
previous studies.
DIB features have also been detected in the commissioning spectra of the
Guoshoujing Telescope (LAMOST) of resolving power similar to that of SDSS.
Detections of DIBs towards hundreds of thousands of stars are expected from the on-going and 
up-coming large scale spectroscopic surveys such as RAVE, SDSS III and LAMOST,
particularly from the LAMOST Digital Sky Survey of the Galactic Anti-center (DSS-GAC).
Such a huge database will provide an unprecedented opportunity to study the
demographical distribution and nature of DIBs 
as well as using DIBs to probe the distribution and properties of the ISM and the dust extinction.}
\end{abstract}

\begin{keywords}
ISM: lines and bands -- ISM: dust, extinction -- stars: general.
\end{keywords}

\section{Introduction}
Diffuse interstellar bands (DIBs) are absorption features ubiquitously detected in the
spectra of reddened stars. They are observed from the near UV, optical, to the near infrared,
and arise from the interstellar medium (ISM).
DIBs were first discovered by Heger (1922).
To date over 400 DIBs have been detected (Sarre 2006; Hobbs et al. 2008, 2009).
DIBs are also detected in extragalactic sources, such as the Small Magellanic Cloud (Ehrenfreund et al. 2002; Welty et al. 2006; Cox et al. 2007),
the Large Magellanic Cloud (Ehrenfreund et al. 2002; Welty et al. 2006; Cox et al. 2006),
the spiral galaxy NGC\,1448 (Sollerman et al. 2005), the Andromeda galaxy M\,31 (Cordiner et al. 2008a, 2011),
the Triangulum galaxy M\,33 (Cordiner et al. 2008b), dusty star-burst galaxies (Heckman \& Lehnert 2000) and
Ca~II and damped Ly$\alpha$ absorbers towards quasars
at cosmological distances (Junkkarinen et al. 2004; York et al. 2006; Ellison et al. 2008; Lawton et al. 2008).

The most prominent DIBs in the optical, e.g., $\lambda$$\lambda$4429, 5780, 5797
\footnote{We cite all DIBs with reference to their nominal air wavelengths.}, are well studied.
They are known empirically to trace the neutral hydrogen phase of the ISM,
with intensities that correlate well with the line-of-sight color excess \ebv, the neutral hydrogen column density and
the Na~I column density (e.g., Herbig 1993; Friedman et al. 2011).
Correlations between different DIBs deduced from samples of spectra towards various sight-lines have been used
to infer the origin(s) of the bands -- sets of well-correlated DIBs probably
arise from the same or similar carriers (Krelowsky \& Walker 1987; Josafatsson \& Snow 1987; Westerlund \& Krelowsky 1989;
Cami et al. 1997; Moutou et al. 1999; Wszolek \& Godlowski 2003; Friedman et al. 2011).
Although DIBs have been widely detected and studied, none of them have been spectroscopically identified
and their carriers (widely believed to be grains or molecules) remain elusive (Herbig 1995; Sarre 2006).

DIBs exhibit a wide range of widths, with full widths at half maximum (FWHM) ranging from below 1\,{\AA} to tens of {\AA}.
High-resolution spectroscopy of reddened OB stars has been used to search for new features and to study the intrinsic profiles,
fine structures and possible relations with other sharp interstellar absorption lines (e.g., Sarre et al. 1995; Galazutdinov et al. 2003;
Thorburn et al. 2003; Hobbs et al. 2008, 2009; McCall et al. 2010; Friedman et al. 2011; Vos et al. 2011).
Luminous OB stars are targeted because their spectra possess few photospheric absorption lines,
making it much easier to detect and measure the relatively weak DIBs. Hitherto, more than one hundred high-resolution spectra of OB stars have been accumulated.
For DIBs of intermediate widths, such as the strong DIBs~$\lambda$$\lambda$5780, 6283,
low-to-intermediate resolution spectroscopy matches their widths very well, and thus can be used to detect and study them efficiently.
At a resolving power of $\sim$~3,000, Cordiner et al. (2008a, 2008b, 2011) detected the DIBs~$\lambda$$\lambda$5780, 6283 and several other features in the
optical spectra of a number of supergiants in M\,33 and M\,31 with the DEIMOS spectrograph of the Keck telescope.
Using the Radial Velocity Experiment (RAVE, Steinmetz et al., 2006; Zwitter et al. 2008; Siebert et al. 2011)
data of a resolving power of $\sim$~7,500, Munari et al. (2008) detected the DIB~$\lambda$8621 towards 68 hot stars.
They found a tight correlation between strengths of the DIB~$\lambda$8621 and reddening, implying that the feature could be a good measure of extinction.
This feature will be targeted by the forth-coming next generation astrometric satellite GAIA at a resolving power of $\sim$~11,500 (Katz et al. 2004).
Using blue-violet spectra at a resolving power of 2,500 from the Galactic O-star spectroscopic survey (GOSSS, Ma{\'{\i}}z Apell{\'a}niz et al. 2011),
Penades Ordaz et al. (2011) detected DIBs~$\lambda$$\lambda$4501,4726,4762,4780,4964 towards a few hundred O stars.

The Sloan Digital Sky Survey (SDSS; York et al. 2000) has obtained low-resolution ($\Delta~\lambda \sim$ 3.3\,{\AA} at 6,000\,{\AA}) spectra of over 500,000 stars
in its Data Release 8 (DR8; Aihara et al. 2011). A fraction of the stars are towards high extinction sight-lines.
In this paper, we report detections of DIBs in the SDSS spectra of a variety of different types of stars.
DIBs~$\lambda$$\lambda$5780, 6283, 
two strong DIB features of intermediate widths that are well matched with the SDSS spectral resolution,
are selected for the investigation. 

The paper is organized as follows.
In Section 2, we introduce the data and method used to identify and measure DIBs.
The results of DIBs~$\lambda$$\lambda$5780, 6283 detected in the SDSS spectra are presented in Section 3.
Detections of DIB features in the commissioning spectra of the Guoshoujing Telescope (LAMOST, Wang et al. 1996) are reported in Section 4.
The prospect and implications of detecting and studying DIBs towards a huge number of sight-lines from several on-going and up-coming large scale low-resolution
spectroscopic surveys are discussed in Section 5.

\section{Data and Method}
We use a template subtraction method to detect DIBs in the SDSS spectra.
This method requires a sample of (reddened) target stars and an (unreddened/low-reddening) control sample
of stars of similar spectral type.
Target stars are defined as stars whose spectra are expected to display DIB features.
Control stars are those whose spectra posses extremely weak or no DIB features,
so that the spectra can be used to construct spectral templates containing no DIBs.
In this Section, selections of target and control stars from the SDSS spectroscopic survey
as well as the technique of template subtraction are introduced.

\subsection{Data}

Both samples of target and control stars are selected from the SDSS Data Release 7 (DR7; Abazajian et al. 2009).
SDSS DR7 contains spectra of more than 300,000 stars.
About 240,000 of them are from the {\rm Sloan Extension for Galactic Understanding and Exploration} ({\rm SEGUE}; Yanny et al. 2009),
focused on the Galactic science. 
We select target stars as those with line-of-sight extinctions \ebv~ $>$ 0.3 and of spectra of signal-to-noise ratios S/N $>$ 40.
We restrict targets to objects with valid determinations of effective temperature \teff, surface gravity \logg~and metallicity \feh~
as given by the SEGUE Stellar Parameter Pipeline
(SSPP, Lee et al. 2008a,b; Allende Prieto et al. 2008; Lee et al. 2011; Smolinski et al. 2011).
Typical uncertainties of those parameters are 180 K, 0.25 dex and 0.23 dex
for \teff, \logg~and [Fe/H], respectively, dominated by systematic errors (Schlesinger et al. 2010; Smolinski et al. 2011).
On the other hand, in the current work we are mainly interested in the relative ranking of stars in the \teff, \logg~and \feh~ parameter spaces,
such that systematic uncertainties are not important.
For robust template subtraction, we further require the errors of \teff, \logg~and \feh~to be smaller than 100 K, 0.2 dex and 0.1 dex, respectively.
Values of \ebv~used here are from the extinction map of Schlegel et al. (1998).
In total, 2,164 target stars are selected. Their spatial and stellar parameter distributions are displayed in Fig.\,1.
The targets come from spectroscopic plates that spread over a large sky area ($-42\degr \le b \le +22 \degr$, ~$8\degr \le l \le 210 \degr$),
and involve stars of a wide range of spectral type, luminosity class and metallicity.
Given the requirements on S/N ratio and line-of-sight extinction,
most targets are from bright SEGUE plates towards low Galactic latitudes.
Essentially, all the target stars are brighter than 17 in $g$ band.
The requirements suggest that the majority of the target stars are from the Galactic disk, with the remaining small fraction from the halo.
This is consistent with the fact that most targets have \feh~higher than $-$0.7,  as shown in the bottom left panel of Fig.\,1.

\begin{figure}
\includegraphics[width=90mm]{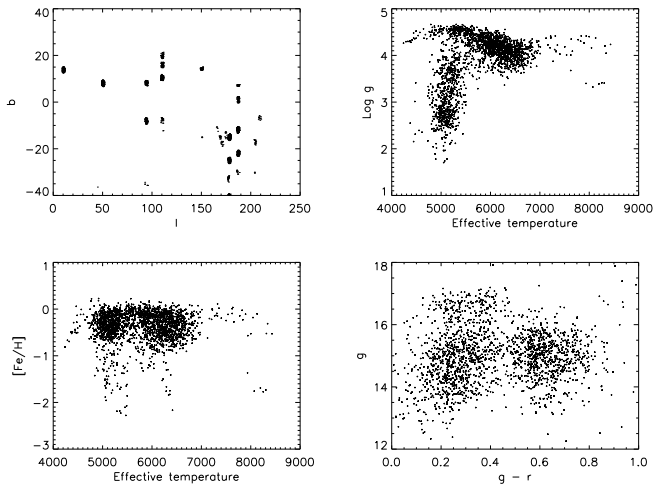}
\caption{Properties of the sample of target stars.
\emph{Top left:} Spatial distribution in Galactic longitude $l$ and latitude $b$.
\emph{Top right:} $\teff$ versus \logg.
\emph{Bottom left:} $\teff$ versus [Fe/H].
\emph{Bottom right:} $g-r$ versus $g$.}
\label{}
\end{figure}

For the control sample, we select stars of S/N ratios $>$ 20 and line-of-sight extinctions \ebv~$<$ 0.03.
The cuts on errors of \teff, \logg~and \feh~are the same as target stars.
In total, 62,266 stars are selected.
Given the large number of control stars, only those with the highest S/N ratios are used
as templates, as described below.

\subsection{Method} The idea of template subtraction method is straightforward.
For each star in the target sample, stars of similar spectral types, luminosity
classes and metallicities are selected from the control sample.
Quantitatively, we require the differences in \teff, \logg~and \feh~between a
control and a target star to be smaller than 100 K, 0.5 dex and 0.2 dex,
respectively.  Secondly, after correcting for differences in radial velocities
of control stars relative to the target star, spectra of the first one hundred
control stars of highest S/N ratios are co-added to construct a template
spectrum of very high S/N ratio.  In cases where the number of control stars is
less than one hundred, all control stars satisfying the above criteria are
used. For the two DIB features of interest here, $\lambda$5780 and
$\lambda$6283, a segment of the template spectrum of a width of 200\,{\AA}
centered around the feature of concern is scaled to the target spectrum,
subtracted and then rectified using the continuum level determined from the
{\it scaled}\, template spectrum.  This certificated residual spectrum is
devoid of stellar absorption features and enhances DIB features in cases of a
good template matching.  To account for the different amount of extinction and
other types of mismatch of the template and target spectra, the scaling factors
of the segment of the template spectrum to target spectrum are fitted with a 
linear polynomial of constant slope after masking the DIB feature.  Note that the strength
of a DIB feature is very sensitive to the scaling factor at the wavelength of
the feature, which is accurately estimated in the above fitting in cases of a
good template matching.  Given the relatively narrow wavelength span under
consideration here, a continuum of constant slope is also used in fitting the
continuum level of the {\it scaled}\, template spectrum.  This continuum level
is assumed to be equally applicable to the target spectrum.  It is much easier
to accurately determine the continuum level of the {\it scaled}\, template
spectrum than that of the target, given the former generally has a very high
S/N ratio that makes it easy to mask and exclude regions of stellar absorption
features in the spectrum.  Finally, we search for the DIB absorption feature
around the expected wavelength in the rectified residual spectrum and measure
its equivalent width using a line fitting method. 

The DIB~$\lambda$5780 contains two blended components, a narrow and a 
much broader one. The narrow and broad components have widths of 2.07 and 15.5\,{\AA} 
centered at 5780.59 and 5779.48\,{\AA}, respectively (Jenniskens
\& D{\'e}sert 1994).  Slightly different values, a width of 2.11\,{\AA} centered
at 5780.48\,{\AA} for the narrow component, were given by Hobbs et al. (2008).
Owing in part to the difficulty in reliably defining the continuum level for
the broad component, in the current work only the strengths of the narrow 
component will be considered, the DIB~$\lambda$5780 refers only to its narrow component.
Given that the typical Doppler spread measured from interstellar 
atomic lines toward reddened stars (\ebv~$<$ 0.8) is about 7.5 \kms 
(Welty \& Hobbs 2001) and the low spectral resolution of SDSS data, 
Doppler broadening is likely to be negligible. As a consequence, using a 
Gaussian  profile of a fixed FWHM 3.85\,{\AA} to fit and measure the strengths of
the DIB~$\lambda$5780 is a good approximation.  
Here 3.85\,{\AA} is the expected FWHM of the DIB~$\lambda$5780  
after taking into account its intrinsic width and the SDSS instrumental 
broadening. 

The DIB~$\lambda$6283 also shows a relatively deep, narrow
feature superimposed on a much shallower, broader component.
Examinations of the DIB~$\lambda$6283 observed with high resolving powers show 
that it exhibits a profile closely resembling a Lorentzian one, with its broad 
wings possibly contaminated by several weak overlapping DIB features 
(Jenniskens \& D{\'e}sert 1993; Drosback 2006). Hobbs et al. (2008) estimated
a FWHM of 4.77\,{\AA} and a central wavelength of 6283.83\,{\AA} for the 
DIB~$\lambda$6283 from the echelle spectra of the double-lined spectroscopic binary 
HD204820. Drosback (2006) obtained a FWHM of $4.0\pm0.3$\,{\AA} from a larger sample. 
In this work, a fixed-width Voigt profile resulting from 
a convolution of a Lorentz profile of a FWHM 4.0\,{\AA} and a Gaussian
profile of a FWHM 3.49\,{\AA} (the instrumental width of SDSS spectra), is 
used to fit and measure the strengths of DIB~$\lambda$6283. 
Note that the DIB~$\lambda$6283 feature is badly affected by telluric absorption features.
However, all released SDSS spectra have been corrected for telluric absorptions by
dividing spectra of hot standard stars, 
so the measurements of DIB~$\lambda$6283 in this work should not be much affected by telluric absorptions.

Sample spectra and the resultant fits are illustrated in Fig.\,2.  Note in this
Figure the spectra are plotted in vacuum wavelengths. 
The standard formula given by Morton (1991) have been
used to convert between air and vacuum wavelengths.

\begin{figure*}
\includegraphics[width=160mm]{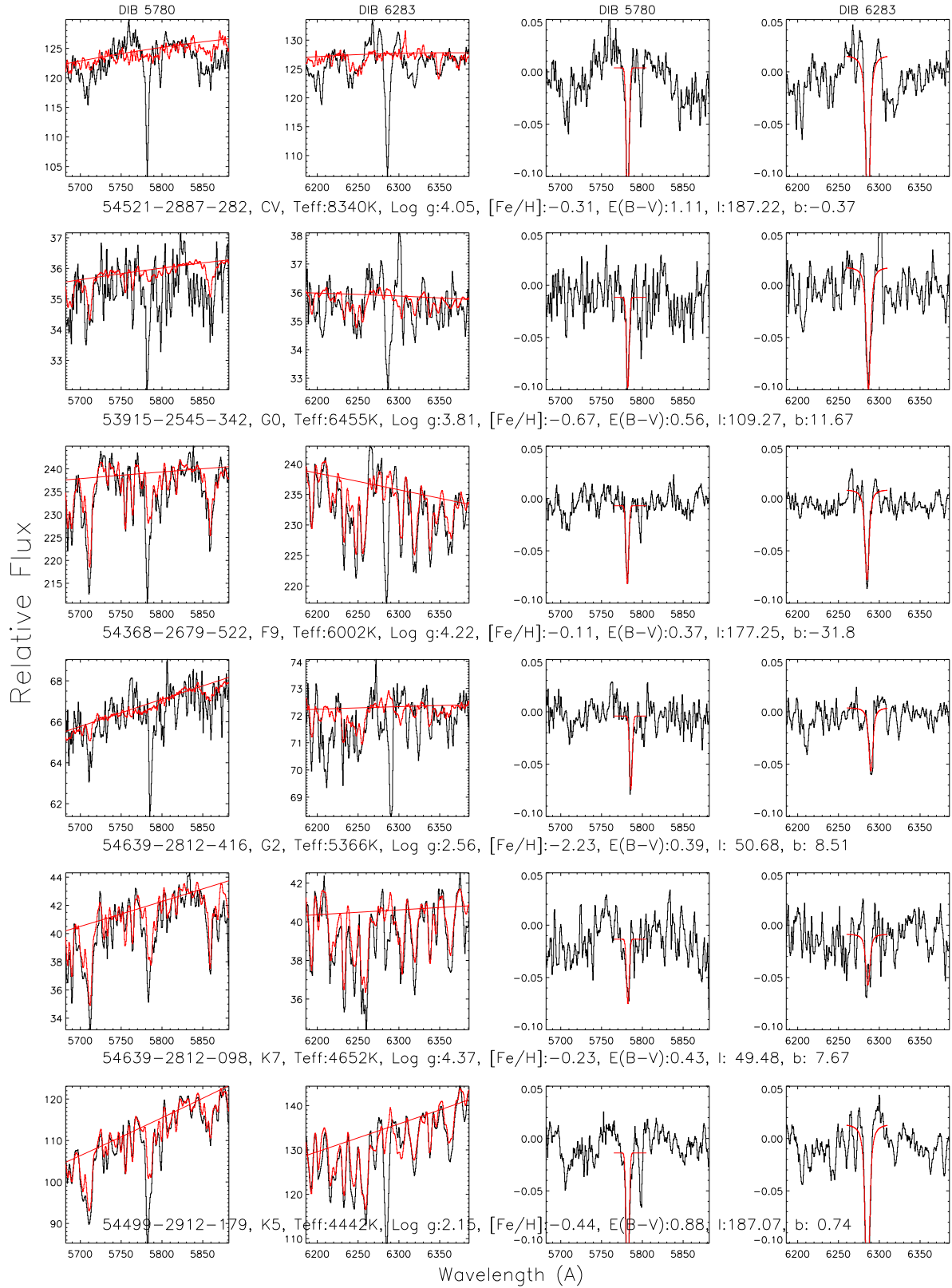}
\caption{The DIB features~$\lambda$$\lambda$5780, 6283 detected in the SDSS spectra of six representative stars. Basic
information of individual stars is listed at the bottom of each row, including, from left to right,
the SDSS spectroscopic target ID, spectral type, effective temperature, surface gravity, metallicity,
line-of-sight extinction, Galactic longitude and latitude.
For each row, segments of the target (black) and the {\it scaled}\, template (red) spectra centered at 5780\,{\AA} and 6300\,{\AA} are shown in the left two panels, respectively.
The lines in red show fitted continuum levels of the {\it scaled}\, template spectra, assumed to be equally applicable to the target spectra.
Those continua were used to rectified the target spectra after subtracting the {\it scaled}\, template ones. The residuals 
are plotted in the right two panels in black, with the Gaussian fitting to the DIB~$\lambda$5780 and Voigt fitting to the DIB~$\lambda$6283 over-plotted in red (see text for detail).
Note that the DIB~$\lambda$5797 is well detected in the spectra of the 1st, 3rd and 6th rows,
and the DIB~$\lambda$6203 is well detected in the spectra of the 1st, 2nd and 4th rows.
}
\label{}
\end{figure*}

\section{Results}

The DIBs~$\lambda$$\lambda$5780, 6283 are detected in spectra of almost all target stars of a wide variety of types, from hot A
to cool K types, dwarfs and giants, metal-rich and -poor stars.
Dozens of target stars which lack obvious DIB features are nearby cool dwarfs and thus suffer little extinction
[The extinction map of Schlegel et al. (1998) used to select target stars refers to the total extinction for a given sight-line].
Fig.\,2 displays DIBs~$\lambda$$\lambda$5780, 6283 detected in SDSS spectra of six representative target stars.
Basic information of individual example stars is listed in the Figure, including, from left to right,
the SDSS spectroscopic target ID, spectral type, effective temperature, surface gravity, metallicity,
line-of-sight extinction, Galactic longitude and latitude.
The stellar parameters are those given by the SSPP pipeline, and extinction values are from Schlegel et al. (1998).
As shown in the spectrum of a K7 dwarf plotted in the 5th row, DIB~$\lambda$5780 suffers serious contamination from
stellar absorption lines in cool and metal-rich stars, leading to some uncertainties in the
identification and measurement of the feature.
To verify the identification of the DIB~$\lambda$5780 feature in the spectrum of this K7 dwarf, we visually inspected individual spectra of its
control stars and found that they agree with each other, confirming that the identification is real.
Note that DIBs~$\lambda$$\lambda$5797,6203 have been detected in more than half of the spectra.
Several other DIB features, such as the DIBs~$\lambda$$\lambda$5705, 5849, 6195, 6269, 6308, 6317, 6376, 6379 possibly have also been
detected in some of the spectra.

For the DIBs~$\lambda$$\lambda$5780, 6283, we have measured their depths and central wavelengths by Gaussian and Voigt profile fitting, respectively. 
The equivalent widths (EW) and radial velocities are also calculated. To derive the radial velocities, the rest-frame wavelengths of
DIBs~$\lambda$$\lambda$5780, 6283 given by Hobbs et al. 2008 are used. Examples of the fitting are over-plotted in Fig.\,2.
In most cases, the target and template spectra match well and the fitting performs reasonably.
In a few cases where there are not enough control stars to construct a matching template of high fidelity,
the DIB features are weak, the spectra are of low S/N ratios or there are some defects in the spectra,
the fitting procedure may deliver poor or even unreliable results.
Nevertheless, we can still confirm the presence of the DIB features in most of those spectra by visual inspection.

$EW$(5780) and $EW$(6283) are very sensitive to the continuum levels, particularly for $EW$(6283) because of its broad width.
To estimate the uncertainties caused by continuum determinations, we have re-fitted the DIBs~$\lambda$$\lambda$5780, 6283 
with their continuum levels fixed to zero. For the DIB~$\lambda$5780, we find that values of $EW$(5780) measured with varying continuum levels 
are strongly correlated with (within an uncertainty of 0.03\,{\AA}), but 10 per cent systematically smaller than those measured with fixed ones.
This is because the broad component of the DIB~$\lambda$5780 will generally make the continuum level of its narrow component below zero when
fitted with varying continuum levels. For the DIB~$\lambda$6283, values of $EW$(6283) measured with varying continuum levels
are also strong correlated with (within an uncertainty of 0.09\,{\AA}), but 18 per cent systematically larger than those measured with fixed ones.
This is because there are plenty of DIBs around the DIB~$\lambda$6283 within 100\,{\AA},
including the relatively strong DIBs~$\lambda$$\lambda$6195, 6203, 6269, 6308, 6317, 6376, 6379. 
They will generally cause a slightly under-estimated continuum level of a target spectrum.
The DIB~$\lambda$6283 feature is badly affected by telluric absorption features. 
All released SDSS spectra have been corrected for telluric absorptions.
To check the uncertainties of SDSS
telluric corrections and their effects on the $EW$(6283) measurements, 
we have selected a number of SDSS high S/N ratio spectra of
low-extinction hot white dwarfs taken under different airmasses (1.0 -- 1.2).
The spectra have S/N ratios greater than 40 and \ebv~ extinctions less than
0.03. Examinations of those spectra show that the uncertainties of EW(6283)
caused by any residuals in the telluric absorption corrections are likely to be
well below 0.1 A. 

Detections of DIBs by Gaussian/Voigt profile fitting may lead to false detections when the spectra are very noisy, or when stellar features
are improperly subtracted.
To avoid false detections, we have excluded a total number of 304 targets that have either $EW$(5780) $<$ 0.082\,{\AA} 
or $EW$(6283) $<$ 0.2\,{\AA} in the following analysis. Although such a cut can not get rid of all false detections, 
the remaining ones, if any, are unlikely to affect the main results of the current work.

\begin{figure}
\includegraphics[width=90mm]{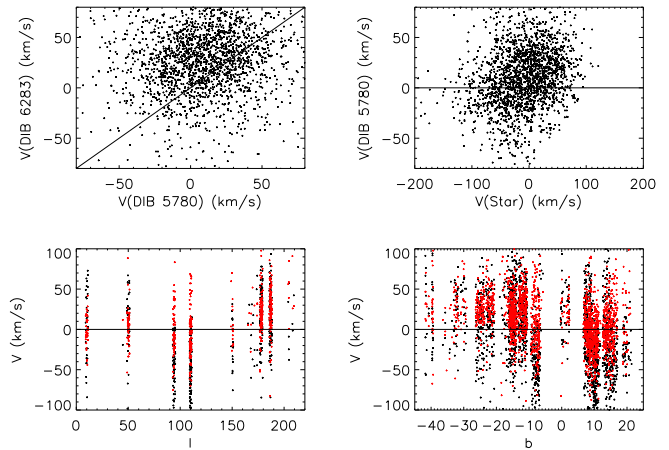}
\caption{\emph{Top left:} Comparison of velocities measured from two DIB features $\lambda$5780 and $\lambda$6283, respectively.
\emph{Top right:} Comparison of stellar and DIB~$\lambda$5780 velocities. 
\emph{Bottom left:} Stellar (black) and DIB~$\lambda$5780 (red) velocities as a function of Galactic longitude.
\emph{Bottom right:} Stellar (black) and DIB~$\lambda$5780 (red) velocities as a function of Galactic latitude.
To avoid crowding, only one in four randomly selected data points are plotted in the bottom left panel.}
\label{}
\end{figure}

\begin{figure*}
\centering
\epsfig{file=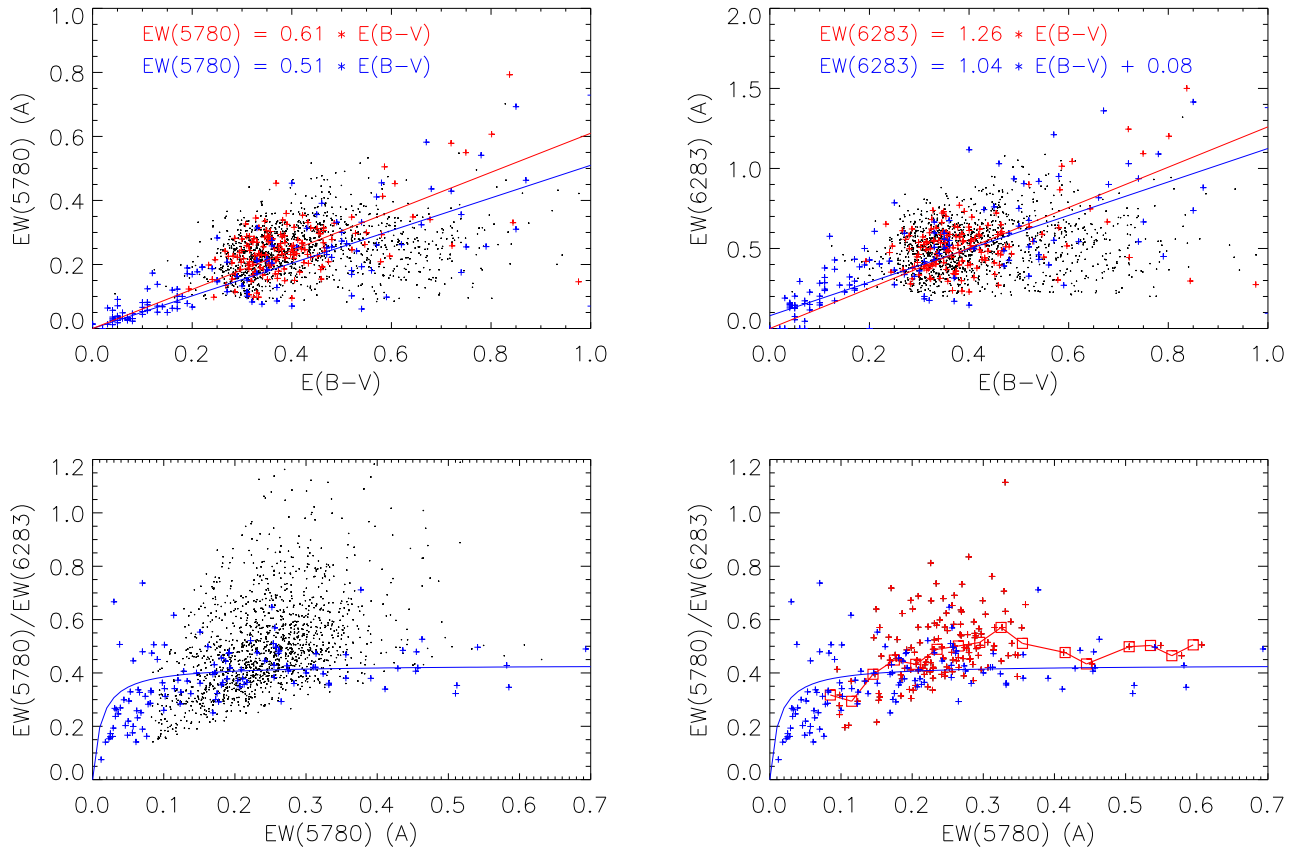,width=18cm}
\caption{ 
\emph{Top left:} $EW$(5780) versus \ebv.
\emph{Top right:} $EW$(6283) versus \ebv.
\emph{Bottom panels:} $EW$(5780)/$EW$(6283) versus $EW$(5780).
The black dots denote the targets of the current work.
The red crosses donote the targets of the current work of spectral S/N ratios higher than 60.
The red lines in the top panels represent regressions of the current work (using targets
of spectral S/N ratios higher than 60).
For comparison, the data and regressions of Friedman et al. (2011) are over-plotted in blue.
In the bottom right panel, the red crosses are divided into 18 bins with a width of 0.03 {\AA},
the red squares in the red line indicate the median value for each bin.
Note that the definitions of $EW$(5780) and $EW$(6283) in Friedman et al. (2011) and this work were very close,
but $EW$(5780) and $EW$(6283) in Friedman et al. were computed with direct integration.
Also note that \ebv~values plotted in the top panels were all derived from the
photometric colors of the target stars.}
\label{}
\end{figure*}

The top left panel of Fig.\,3 compares radial velocities deduced from the DIBs~$\lambda$5780 and $\lambda$6283, respectively.
It can be seen that the velocities deduced from DIB~$\lambda$6283 are 16.0 \kms systematically larger than those deduced from DIB~$\lambda$5780,
which is likely caused by uncertainties of the adopted DIB rest-frame wavelengths from Hobbs et al. (2008). After accounting for the systematic differences,
the two features give consistent results within an uncertainty of 32 \kms.
Given that the DIB~$\lambda$5780 feature is narrower and more symmetric than the $\lambda$6283 feature,
the wavelengths of the DIB~$\lambda$5780 in the SDSS spectra are more easily to determine as well as its rest wavelength, 
so velocities deduced from the DIB~$\lambda$5780 are used to represent the DIB velocities. 
The top right panel compares velocities deduced from the DIBs~$\lambda$5780 with velocities of the target stars from the SSPP pipeline.
A weak correlation between the DIB and stellar velocities is seen. The correlation is 
due to the fact that most targets are disk stars and thus
have kinematic characteristics similar to the absorbing interstellar clouds from which the DIBs originate.
Also note compared to the wide range of stellar velocities, the span of the DIB velocities is much limited.
This is expected considering the absorbing clouds are confined to the Galactic thin disk and have more circular orbits than stars.
That the DIB velocities follow the stellar values but with smaller scatters at a given longitude or latitude is
clearly seen in the lower two panels of Fig.\,3. Again, the trends are consistent with the fact
that DIBs arise from the interstellar clouds in the Galactic thin disk.

It is well known that to some extent strengths of individual DIBs are correlated with each other,
as well as the amount of intervening dust extinction.
Using measurements from over one hundred stellar spectra of high S/N ratios and spectral resolution ($R \sim$~38,000),
Friedman et al. (2011) find that $EW$(5780) $= 0.505~\times$~\ebv, $EW$(6283) = 1.045~$\times$~\ebv + 0.080 and
$EW$(5780) $= 0.431~\times$~$EW(6283) - 0.012$, with a correlation coefficient of 0.82, 0.82 and 0.96, respectively.
Note that the definitions of $EW$(5780) and $EW$(6283) in Friedman et al. (2011) and this work were very close, 
but $EW$(5780) and $EW$(6283) in Friedman et al. were computed with direct integration.
Note also values of \ebv~of Friedman et al. (2011) were estimated from the observed and intrinsic colors.
To investigate relations between $EW$(5780), $EW$(6283) and \ebv~using
measurements of stars in the target sample of the current work, the largest ever assembled hitherto,
we plot $EW(5780)$ and $EW(6283)$ as a function of \ebv~ and $EW(5870)/EW(6283)$ against $EW(5780)$ in Fig.\,4.
For comparison, data and relations from Friedman et al. (2011) are over-plotted.

Given that values of \ebv~from the extinction map of Schlegel et al. (1998) suffer from the following limitations:
a) The map gives the total amount of extinction integrated along a given line-of-sight to infinite, and consequently the value is an upper limit of the
real one for a local disk star; b) The map fails at low Galactic latitudes ($|b| \le 5\degr$);
and c) The map has a limited spatial resolution about 6 arcmin, whereas the extinction may well vary below such a scale,
we did not use the values of \ebv~from Schlegel et al. in Fig.\,4.
The \ebv~value of a target used here was derived directly from the difference of its observed and intrinsic $g - r$ colors.
The intrinsic $g - r$ color was chosen to be the median value of the dereddened $g - r$ colors of its control stars.
Here $A_g/E_{B-V} = 3.303$ and $A_r/E_{B-V} = 2.285$ at a total-to-selective extinction ratio of 3.1 were used (Schlafly \& Finkbeiner 2011).

The variations of $EW$(5780) and $EW$(6283) as a function of \ebv~deduced from
the current work are shown in the top left and right panels of Fig.\,4,
respectively.
The black dots denote the targets of the current work.
The red crosses denote the targets of the current work of spectral S/N ratios higher than 60.
Linear regressions passing through the origin were performed on the red crosses in the top panels, 
with each data point carrying equal weight. The relations obtained, i.e., $EW$(5780) $= 0.61~\times$~\ebv~ 
and $EW$(6283) = 1.26~$\times$~\ebv, are plotted in red lines.
The blue crosses and lines denote the data of and the relations obtained by Friedman et al. (2011), respectively.
The slopes of regressions of this work are about 20 per cent larger than those of Friedman et al. (2011).
However, considering the large scatters in both data, this work yields consistent results with Friedman et al. (2011).

Values of the $EW$(5780)/$EW$(6283) ratios as deduced for the current target sample stars are plotted against $EW$(5780) 
in the bottom panels of Fig.\,4. 
The symbols are the same as in the top panels.
In the bottom right panel, red crosses are divided into 18 bins with a width of 0.03 {\AA},
red squares in the red line indicate the median value for each bin. 
The data in this work yield a median value of 0.48,  consistent with the result of $\sim$~0.42 of Friedman et al. (2011).
The data show large scatters, possibly due to the limited S/N ratios of the target spectra. 
The scatters of data deduced from spectra of S/N ratios higher than 60 are comparable to those of Friedman et al. (2011), 
implying measurements of the current work are robust. 
The red line in the bottom right panel seems to indicate that $EW$(5780)/$EW$(6283) first increases with $EW$(5780) and then saturates for 
$EW$(5780) above some value between 0.2 and 0.3 {\AA}.
This interesting trend is supported by the data of Friedman et al. (2011).
Friedman et al. attribute the behaviors of the strength of the DIB~$\lambda$5780 relative to that of 
$\lambda$6283 to measurement errors or a threshold effect, where one DIB cannot form until a significant amount of another DIB is present.
Additionally, it could be interpreted that the fractions of carrier of individual DIBs depend on certain local environment factor(s) in an equilibrium state.
For example, Vos et al. (2011) find that the $EW$(5780)/$EW$(5797) ratio depends on cloud density
as well as exposure to the interstellar radiation field.

Taking into account the uncertainties of $EW$(5780), $EW$(6283) and \ebv,
the results plotted in Fig.\,4 are consistent with the data 
obtained by Friedman et al. (2011),
particularly for those deduced from spectra of S/N ratios higher than 60.
Both the data of the current work and Friedman et al. (2011) show large scatters,
suggesting that a significant fraction of the scatters are intrinsic.
We will investigate whether the properties of DIBs vary with Galactic position and environment in a separate paper,
using \ebv~values and distances of stars determined directly from photometric and spectroscopic observations.

\section{DIBs in the LAMOST commissioning spectra}
\begin{figure*}
\includegraphics[width=180mm]{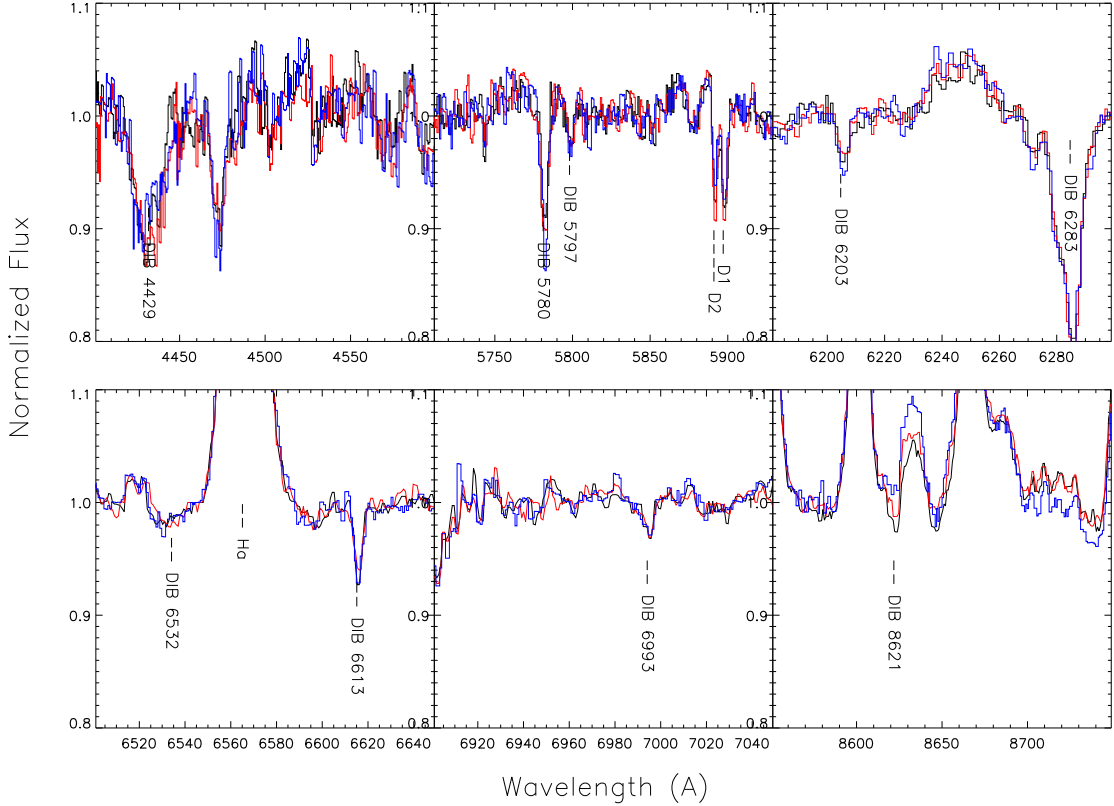}
\caption{
Nine DIB features have been detected in the LAMOST commissioning spectra of an emission line star. The spectra have been normalized.
The different colors in each panel refer to individual spectra from three separate exposures.
}
\label{}
\end{figure*}

The Guoshoujing Telescope (the Large Sky Area Multi-Object Fiber Spectroscopic Telescope LAMOST; Wang et al. 1996; Su et al.
1998; Xing et al. 1998; Zhao 2000; Cui et al. 2004; Zhu et al. 2006; Cui et
al. 2012; c.f. http://www.lamost.org/website/en/), is an
innovative quasi-meridian reflecting Schmidt Telescope that is capable of
recording spectra of up to 4,000 objects simultaneously in a field of view of
5$^\circ$ in diameter. 
Following a two-year commissioning initiated in September, 2009 and a one-year pilot survey initiated in October, 2011,
the LAMOST formal Galactic and extragalactic surveys are expected to commence in fall of 2012 (Zhao et al. 2012).

Given the similar spectral resolving powers of SDSS and LAMOST,
DIBs should be also detectable in the LAMOST spectra.
Given that the data reduction pipelines of LAMOST are still under development and reliable stellar parameter estimates are not available yet,
the template subtraction method currently does not work for the LAMOST data.
However, taking advantage of the smooth stellar continua of hot early-type stars,
DIBs are also easily identified in the LAMOST low-resolution spectra.
As an example, at least nine DIB features ($\lambda$$\lambda$4429, 5780, 5797, 6203,
6283, 6532, 6613, 6993, 8621) have been detected in the spectra of a LAMOST target (Fig.\,5). 
Note the prominence of DIBs~$\lambda$$\lambda$5780, 6283.
The target was observed on December 21, 2009 with the half-slit (1.65 arcsec) mode (R $\sim$~2,000) during the commissioning phase of LAMOST.
Three exposures, each of 900s, were obtained. The data were reduced using the LAMOST 2D pipeline
(Luo, Zhang \& Zhao 2004) following the standard procedure, including bias subtraction, cosmic-ray
removal, 1D spectral extraction, flat-fielding and wavelength calibration. Sky subtraction
was not performed. This did not affect the identification of DIB features given the weakness of sky continuum level compared to that of the target.
Flux calibration was also not necessary.
It is worth mentioning that the target is an emission line star first discovered by Kohoutek \&~Wehmeyer (1999).
Its broad and prominent  H${\alpha}$ emission line is clearly seen in the bottom left panel of Fig.\,5.

\section{Discussion}

Using a template subtraction method, we have successfully identified the DIBs~$\lambda$$\lambda$5780, 6283 in the SDSS low-resolution spectra of
a sample of about 2,000 stars and measured their equivalent widths and velocities.
The sample is by far the largest of stars with DIBs detected and measured.
The sample stars are distributed over a large sky area and cover a wide range of stellar parameters
of effective temperature, surface gravity and metallicity, confirming that the carriers of DIBs are ubiquitous in the diffuse ISM.
We use the sample to investigate the relations between the equivalent widths of the DIBs~$\lambda$$\lambda$5780, 6283 and sight-line extinctions,
and find results consistent with previous works. The data however show large scatters.
Some of the scatters are due to the measurement uncertainties,
especially for spectra of limited S/N ratios.
A significant fraction of the scatters are probably intrinsic -- the data of Friedman et al. (2011) also show significant scattering.
Identifying potential environmental factors that may lead to the complexities of observed properties of DIBs is pivotal to reveal the nature of DIBs.
To achieve such a goal, accurate measurements of DIBs towards a large number of sight-lines probing a variety of environments with low-cost 
spectra of low and intermediate resolution as illustrated in the current work should be an effective approach.

Apart from the limited S/N ratios of spectra, 
other sources of uncertainties in measuring the DIBs~$\lambda$$\lambda$5780, 6283 in this work mainly include a 
mismatch between a target and template spectrum, continuum determinations 
and profile fitting of the DIB features with a single Gaussian/Voigt.
A mismatch of a target and template spectrum may occur if the target is of a rare spectral type, 
the stellar parameters of the target star are in significant error or there are some defects (e.g., residual telluric absorption features) in its spectrum.
Such cases can in principle be flagged out and excluded from the sample in the future.
A mismatch can also happen if the distribution of stellar parameters of the control sample is biased.
Such cases can be avoided by more careful selections of control stars.
To reduce the uncertainties of continuum levels, sophisticated ways to define stellar continuum are needed.
To reduce the uncertainties introduced by a single Gaussian/Voigt profile fitting,
one can explore deblending the DIB~$\lambda$5780 with two Gaussians and fitting the DIB~$\lambda$6283 with measured profiles.
It might be possible to carry out the fitting with the central wavelengths of individual components of the DIBs tied together, 
reducing the number of degrees of freedom of the fitting function. For spectra of high S/N ratios, 
it might be feasible to define indexes of DIBs and measure them by direct integration.

It is worth mentioning that hot stars have few strong absorption lines blending with DIBs.
In such cases, template spectra are not even needed to identify DIBs.
As illustrated by the example presented in the previous Section, in addition to the strong DIBs~$\lambda$$\lambda$5780, 6283, seven other DIBs can be 
easily identified in low-resolution spectra of such a hot star. The strengths of the DIB features can also be measured by direct integration
given sufficient high S/N ratios. Given that the light of a star-burst galaxy is dominated by contribution from luminous hot stars,
searching for the strong DIBs (e.g., DIBs~$\lambda$$\lambda$5780, 6283) in their SDSS spectra might be feasible.

Searching and measuring DIBs in low-resolution spectra, now widely available for a huge number of objects,
have obviously the potential to tackle a number of problems of the ISM. For example, they can be used to 
probe the distribution and properties of the DIBs, estimate the extinction towards individual stars, trace the amount of ISM 
absorption and search for extremely metal-poor stars. The data will help unveil the nature of DIBs.
Large scale spectroscopic surveys, such as the on-going RAVE and SDSS, and the up-coming LAMOST Galactic spectral surveys
will yield detections of DIBs (e.g., $\lambda$$\lambda$4429, 5780, 5797, 6203, 6283, 6613, 6993, 8621)
in hundreds of thousands stars of a variety of spectral types, located in different parts and environments of the Milky Way.
The huge database will provide an unprecedented opportunity to study the demographical distribution and nature of DIBs
as well as using DIBs to probe the distribution and properties of the ISM and the dust extinction.
Given the large sample, targets exhibiting particularly interesting features can be selected and followed 
with high-resolution spectroscopic observations.

Strengths of DIBs measured with low-resolution spectroscopy can be used to estimate extinction towards a large number 
of individual stars targeted in large scale spectroscopic surveys.
Extinction is one of the most fundamental stellar parameters and affects the determinations
of others (i.e. effective temperature, metallicity and distance). As mentioned above,
extinction values given by Schlegel et al. (1998) have several limitations, especially for the low Galactic latitude regions.
From the relations between the strengths of DIBs and the line-of-sight color excess, strengths of DIBs measured with low-resolution spectroscopy
provide an independent way to estimate extinction of individual stars.
This is especially useful for the up-coming LAMOST Digital Sky Survey of the Galactic Anti-center (DSS-GAC; Liu et al. 2012, in prep).
The DSS-GAC project aims to obtain spectra for a statistically complete sample of over three million stars down to 18.5 limiting magnitude in r- band,
distributed in a contiguous area of 3438 square degree ($-30\degr \le b \le +30 \degr,~150 \degr \le l \le 210 \degr$) 
and sampling a significant volume of the Galactic thin/thick disks, halo and their interfaces.
The project will target approximately 1,000 stars per square degree outside the Galactic plane ($|b|$ $\ge$ 3.5$\degr$) and twice that within the Galactic plane.
Although extinction towards Galactic anti-center is much lower than towards the center, there are regions of high extinction where the extinction map of
Schlegel et al. (1998) is inapplicable. On the other hand, information of extinction is required
to select spectral standard stars and perform flux calibration. Information of extinction is also required to 
interpret the target selection algorithms of the survey and derive basic stellar parameters and the selection function of the survey.
The possibility of estimating the extinction to individual star via the observed DIB strengths, 
unaffected by flux calibration, should be very helpful to the DSS-GAC project.
In the meantime, the DSS-GAC project will deliver a huge database to study DIBs. The database will generate a
three-dimensional extinction map towards the Galactic anti-center, with distances of individual stars estimated
from either a spectro-photometric analysis or directly from the parallax measurements to be delivered by the forth-coming 
GAIA mission (Perryman et al. 2001). Finally, given the tight correlations between DIB strengths (especially the DIB~$\lambda$5780; Friedman et al. 2011)
and the neutral hydrogen column density, the database can be further used to derive the three-dimensional distribution of the neutral hydrogen,
probe the properties, structure and kinematics of the ISM, and constrain the peculiar motion of the sun with respect to the Local Standard of Rest.

It is possible that detections of DIBs with low-resolution spectroscopy may provide a way to correct for contamination of the ISM Ca~II absorption
when searching for extremely metal-poor stars with low-resolution spectroscopic surveys.
Discovery of extremely metal-poor stars is of great importance in understanding the nature and the initial mass function of the first stars,
the processes of element production during the early evolution phase of the Milky Way and the metallicity distribution function of the Galactic
halo (Beers \& Christlieb, 2005). Low-resolution spectroscopic surveys have been playing an important role in
searching for very metal-poor stars (Beers \& Carollo 2008; Beers et al 2009; Fulbright et al. 2010).
Under low spectral resolution, the stellar Ca~II absorption lines blend with absorption arising from 
intervening ISM, leading to overestimated stellar metallicities.
This problem becomes more severe as stars get more metal-poor.
Strengths of certain DIBs might provide a way to correct for the ISM Ca~II absorption.
Using high-resolution spectra of a sample of 35 stars, Galazutdinov et al. (2004) show that the strengths of some
DIBs correlate well with the interstellar K~I lines but to a much lower degree with the Ca~II lines.
From a sample of several hundred low-resolution spectra of O-type stars, Penades Ordaz et al. (2011) find a moderately
low correlation coefficient between the Ca~II~K line absorption and the extinction with a peculiar spatial distribution that
they ascribe to a relationship between absorption saturation and velocity profile of the Ca~II~K line.
Nevertheless, this is a problem worth further investigations.

Finally, we would like to point out that the template subtraction technique used in the current work 
can be adapted to tackle other problems, such as searching for peculiar stars of unusual characteristics 
in a large spectroscopic database where templates of "normal" stars can be easily constructed.
Examples of such peculiar objects include binaries, emission line stars and stars of peculiar chemical composition.

\vspace{7mm} \noindent {\bf Acknowledgments}{
We would like to thank the referee, Dr. Martin Cordiner, for the valuable comments and suggestions, which helped improve the quality of the paper.
We appreciate useful discussions with Prof. Jingyao Huo and Dr. Aigen Li.
We also would like to acknowledge Dr. Yong Zhang for his critical reading and 
Dr. Thomas M. Hughes for improving the writing of the paper.
This work has made use of the Sloan Digital Sky Survey (SDSS) and SIMBAD databases.
Guoshoujing Telescope (the Large Sky Area Multi-Object Fiber Spectroscopic Telescope LAMOST) is a
National Major Scientific Project built by the Chinese Academy of Sciences. Funding for the project has been
provided by the National Development and Reform Commission. LAMOST is operated and managed by the National
Astronomical Observatories, Chinese Academy of Sciences.
}

\label{lastpage}

\end{document}